# Empirical Double-Hybrid Density Functional Theory: A 'Third Way' in Between WFT and DFT

Jan M. L. Martin*[a] and Golokesh Santra[a]

*Dedication: In fond, appreciative memory of Walter Thiel (1947–2019)*

**Abstract:** Double hybrid density functional theory arguably sits on the seamline between wavefunction methods and DFT: it represents a special case of Rung 5 on the "Jacob's Ladder" of John P. Perdew. For large and chemically diverse benchmarks such as GMTKN55, empirical double hybrid functionals with dispersion corrections can achieve accuracies approaching wavefunction methods at a cost not greatly dissimilar to hybrid DFT approaches, provided RI-MP2 and/or another MP2 acceleration techniques are available in the electronic structure code. Only a half-dozen or fewer empirical parameters are required. For vibrational frequencies, accuracies intermediate between CCSD and CCSD(T) can be achieved, and performance for other properties is encouraging as well. Organometallic reactions can likewise be treated well, provided static correlation is not too strong. Further prospects are discussed, including range-separated and RPA-based approaches.

## 1. Introduction

Wavefunction-based ab initio theory (WFT) and density functional theory (DFT) are the two primary approaches of electronic structure. WFT methods can approach exact solutions of the Schrödinger equation to almost arbitrary accuracy (see, e.g., Refs. [1–4] and references therein). Alas, its unfavorable CPU time scaling with system size ($N^7$ for CCSD(T),[5,6] steeper for more rigorous methods) limits its applicability (as of 2019) to small molecules. (At the CCSD(T) level, localized pair natural orbital approaches[7,8] represent an emerging remedy.)

In contrast, density functional theory (DFT) features relatively gentle system size scaling, at the expense of introducing an unknown (and perhaps unknowable) exchange-correlation functional. The proliferation of approximate exchange-correlation functionals has led to what Perdew has termed[9] "the functional zoo".[9,10]

WFT has a well-defined path for convergence to the "right answer for the right reason". While no equivalent exists for DFT, in 2001 Perdew[11] introduced the organizing principle he called "Jacob's Ladder", using a Biblical metaphor (Genesis 28:10–12). Its definition is illustrated in Figure 1, together with example functionals on each "rung".

Earth is the Hartree "Vale of Tears", with neither exchange nor correlation. The Hartree equations can be solved exactly within the finite basis given (or on the real-space grid given),[12] without resorting to any adjustable parameters.

Heaven is the elusive goal of the exact exchange-correlation (XC) functional, and thus the exact solution of the Schrödinger equation.

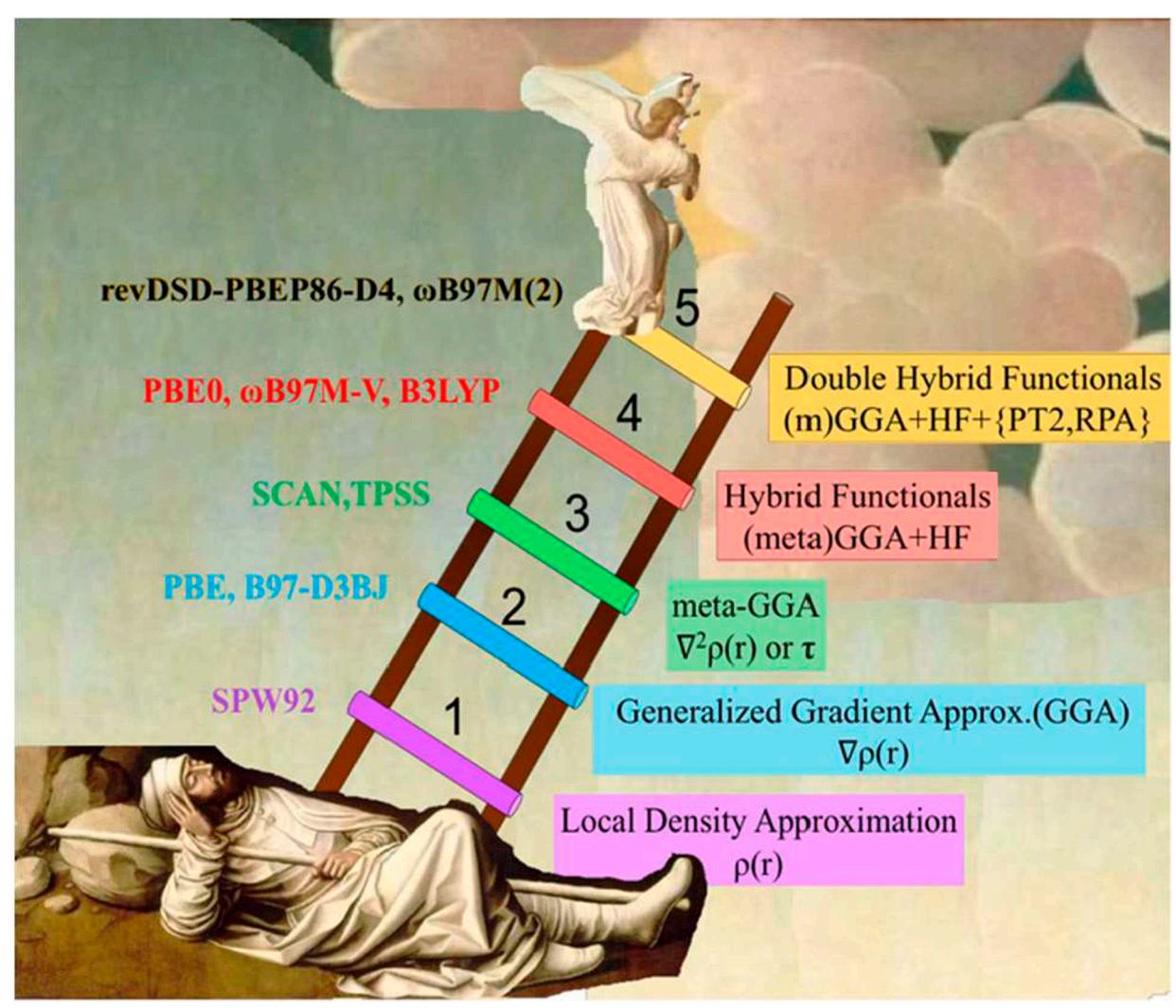

**Figure 1.** Illustration of Jacob's Ladder of DFT.

The local density approximation constitutes rung one: it represents the exact solution for a uniform electron gas of a given density. The exchange energy can be determined analytically (Slater[13]), while different parametrizations[14–17] of the LDA correlation functional represent different fits to the quantum Monte Carlo results[18] for uniform electron gases. The LDA XC functional only depends on the electronic density and not on any derivatives nor any other entities.

[a] *Prof. Dr. J. M. L. Martin, G. Santra*
Department of Organic Chemistry, Weizmann Institute of Science, 76100 Rehovot, Israel
phone: +972 8 9342533
fax: +972 8 9343029
E-mail: gershom@weizmann.ac.il

Rung two corresponds to GGAs or generalized gradient approximations, in which the reduced density gradient is introduced in the XC functional. Examples are the popular BP86[19,20] and PBE[21] functionals.

Rung three consists of meta-GGAs (mGGAs), which additionally involve higher density derivatives (or the kinetic energy density, which contains similar information to the density Laplacian). TPSS[22] is a popular meta-GGA, as is the recent SCAN (strongly constrained and appropriate normed[23]).

Rungs two and three are often collectively referred to as semi-local functionals.

Orbital-dependent DFT[24] covers rungs four and five: on rung four, only occupied orbital-dependency is introduced, while on rung five the unoccupied orbitals make their appearance. The most important rung four functionals are the hybrids, which can be further subdivided into four subclasses:

- global hybrid GGAs, such as the popular B3LYP[25,26] and PBE0[27] hybrids, as well as B97-1.[28] (We note that Hartree-Fock theory itself is a special case, with 100 % exact exchange and null correlation.)
- global hybrid meta-GGAs, such as M06,[29] M06-2X,[29] and BMK[30]
- range-separated hybrid GGAs, such as CAM-B3LYP[31] and ωB97X-V[32]
- range-separated hybrid meta-GGAs, such as ωB97M-V[33]

This leaves rung five, of which we will presently consider one subcase, the double hybrids. More general reviews have been published earlier;[34–36] the present review will focus primarily on our own work, as well as the broader context.

## 2. The GMTKN55 Benchmark

There is no shortage of papers that advocate functional X for property Y of molecule family Z, based on data of greater or lesser quality. In order to make broader statements, however, large and chemically diverse benchmarks are in order. The two largest ones presently available are the gargantuan (4,985 unique data points) MGCDB84 (Main Group Chemistry Data Base, 84 subsets) of Mardirossian and Head-Gordon,[37] and the still very large GMTKN55 (General Main-group Thermochemistry, Kinetics, and Noncovalent interactions database, 55 subsets) of Grimme, Goerigk, and coworkers,[38] itself an expansion and update of earlier GMTKN24[39] and GMTKN30[40,41] databases. GMTKN55 is the benchmark we shall use and discuss throughout the present paper, even though it was developed more than a decade after the oldest DH functional discussed here.

In all, GMTKN55 consists of almost 1500 unique energy differences spread over 55 different problem sets, entailing 2,459 unique single-point energy calculations. The problem sets can be grouped into five major classes: thermochemistry, barrier heights, large molecule reactions, intermolecular interactions, and conformer energies (mostly driven by intra-molecular interactions).

The original GMTKN55 paper prescribes def2-QZVP basis sets for all calculations, with (for rung 5 and WFT methods) all subvalence electrons correlated. The statistic reported is WTMAD2 (weighted mean absolute deviation, type 2), which is defined as:

$$\mathrm{WTMAD2} = \frac{1}{\sum_i^{55} \mathrm{N}_i} \cdot \sum_{\mathrm{i}}^{55} \mathrm{N}_i \cdot \frac{56.84\ \mathrm{kcal/mol}}{\overline{|\Delta \mathrm{E}|}_i} \cdot \mathrm{MAD}_i$$

in which $|\Delta E|_i$ is the mean absolute value of all the reference energies for subset $i$, $N_i$ the number of systems in the subset,

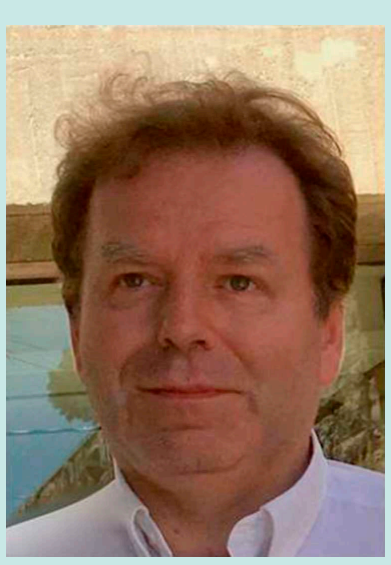

Jan M.L. (Gershom) Martin was born in Belgium in 1964 and obtained his licentiaat (integrated B.Sc./M.Sc. degree) in chemistry in 1987, his Ph.D. in chemistry in 1991, and his Aggregaat voor het Hoger Onderwijs (Habilitation) in 1994, all from U. of Antwerp. From 1987 until 1996 he worked for the Nationaal Fonds voor Wetenschappelijk Onderzoek, first as a doctoral assistant, then as a postdoctoral researcher, finally as a tenured research fellow. Subsequently, he joined the faculty of the Weizmann Institute of Science and was promoted to full professor in 2005, succeeding Meir Lahav as the Baroness Thatcher Professor of Chemistry. He is a Foreign Member of the Royal Academy of Sciences, Literature, and Arts of Belgium and was awarded the 2017 Israel Chemical Society Prize for Excellence, as well as the 2004 Dirac Medal of WATOC.

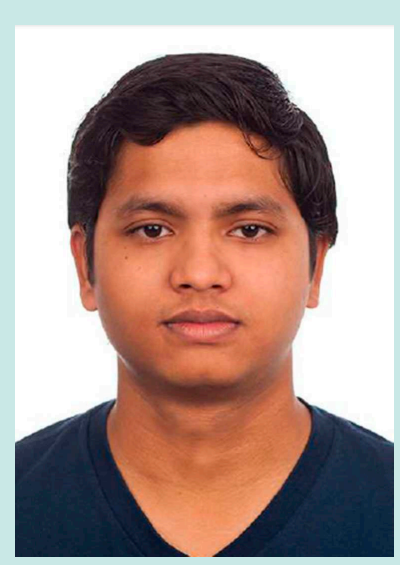

Golokesh Santra was born in India in 1995 and obtained his BSc(Hons.) Chemistry degree in 2015 from Ramakrishna Mission Vidyamandira Belur Math, Howrah, India [an autonomous college under the auspices of U. of Calcutta]. He obtained his MSc degree in Chemistry in 2017 with the academic excellence award from Indian Institute of Technology Kanpur, India. He stayed on for an additional six months as a research associate in Prof Amalendu Chandra's group, followed by a four-month stint with Dr. Marialore Sulpizi at the University of Mainz, Germany. Since August 2018, he has been a Ph.D. student in Prof. Martin's group at the Weizmann Institute of Science.

and $MAD_i$ represents the mean absolute difference between calculated and reference reaction energies for subset *i*. The normalization by $|\Delta E|_i$ compensates for the different energy scales of different types of reaction energies: thus, errors in noncovalent interactions or conformer energies do no "drown in the noise" of, e. g., total atomization energies and ionization potentials that are 2–3 orders of magnitude larger.

We note in passing that MAD is a more "robust statistic"[42] than the root mean square deviation (RMSD), in the statistical sense that MAD is less prone to hypersensitivity to one or a few "outlier" points than RMSD, while the latter (for the same reason) is more useful for spotting "troublemakers". For an unbiased normal distribution, $4\mathrm{RMSD}/5\mathrm{MAD}=(25/8\pi)^{1/2}\approx 1$.[43,44]

In the present work (Table 1) and our recent papers,[45,53] we have adopted GMTKN55 with the following computational protocol modifications:

- Subvalence electrons are frozen according to ORCA's "chemical cores" prescription, i. e., (n-1)*spd* shells of transition metals and heavy *p*-block elements are unfrozen
- For the large-molecule, small-weight subsets C60ISO and UPU23, we employ the def2-TZVPP basis set for computational cost reasons.
- For the remainder, basis sets are identical to or larger than in the original GMTKN55 paper. Specifically, we use def2-QZVPP except that:
- We augment it with diffuse functions, to def2-QZVPPD, for the subsets WATER27, RG18, IL16, G21EA, and AHB21
- The SG-3 integration grid,[66] i. e., a pruned (99,590) grid roughly comparable to the UltraFine grid in the Gaussian program system,[67] is used for GGAs and hybrid GGAs, while for pure and hybrid mGGAs we employed the still larger unpruned (150, 974) grid.

Several authors, notably Gould[68] for GMTKN55 and Chan[69] for MGCDB82, have tried to put said data sets "on a diet", i. e., to reduce them to weighted averages of results for small statistically significant subsamples. All WTMAD2 statistics reported in Table 1 refer to the full GMTKN55 set, however. In addition, we report the partitioning of WTMAD2 among five subsets: basic thermochemistry, barrier heights, large-molecule reactions, intermolecular reactions, and conformer energies (mostly driven by intramolecular noncovalent interactions/ Finally, in the last column we report results for a very recent transition metal benchmark, namely MOBH35 (metal-organic barrier heights, 35 reactions) by Iron and Janes.[65]

## 3. Empirical Double Hybrids

### 3.1 Simple Double Hybrids

While semilocal correlation functionals typically cope quite well with short-range correlation effects, their semilocality makes them intrinsically less suitable for long-range correlation effects, be they dispersion interactions or static correlation.[70] In the early 1990s, Görling and Levy[71] proposed a perturbation theory expansion in a basis of Kohn-Sham orbitals. (We should stress here that, while it is *analogous* to the commonly used Møller-Plesset perturbation theory in a basis of HF orbitals, GLPT2 is by no means *equivalent* to MP2.)

Conceptually, double-hybrid DFT seeks to, on the one hand, combine the strongest features of semilocal correlation and GLPT2, and on the other hand to conjoin the strongest features of semilocal and "exact" exchange through a (global or distance-dependent) linear combination of these two.

In their most general form (omitting range separation parameters for now, for the sake of clarity) double hybrid DFT functionals can be formulated as follows:

(a) In <u>step one</u>, a set of KS orbitals are obtained by solving the KS equations for the following energy functional:

$$E = E_{1e} + c'_X E_{X,HF} + (1 - c'_X) E_{X,DFT} + c'_{c,XC} E_{C,DFT}$$

where $E_{1e}$ is the one electron energy term; $c'_X$ and $c'_{c,XC}$ are the percentage of HF exchange and DFT correlation used, respectively; $E_{X,HF}$ is the HF-like exchange energy, and $E_{X,DFT}$ and $E_{C,DFT}$ are DFT exchange and correlation terms.

(b) In <u>step two</u>, the converged KS orbitals from step 1 are applied to evaluate the energy from the following expression:

$$E_{DH}=E_{1e}+ c_X E_{X,HF}+(1-c_X)E_{X,DFT}+c_{C,XC}E_{C,DFT} + c_{p,ab}E^{postHF}_{\alpha\beta}+ c_{p,ss}\left(E^{postHF}_{\alpha\alpha}+E^{postHF}_{\beta\beta}\right) + E_{disp}[s_6,s_8,a_1,a_2,a_{ABC}, \ldots]$$

in which $c_X$ and $c_{c,XC}$ are the percentages of HF exchange and DFT correlation for this step, $E^{postHF}_{\alpha\beta}$ is the post-HF (MP2, dRPA,…) opposite-spin correlation energy component, while $E^{postHF}_{\alpha\alpha}+E^{postHF}_{\beta\beta}$ is its same-spin counterpart and $c_{p,ss}$ and $c_{p,ab}$ are their coefficients, and $E_{disp}$ is an optional dispersion correction (dependent upon various parameters $s_6, s_8, a_1, a_2, a_{ABC}$ etc. depending on which dispersion model is being employed).

A compact notation for the above is:

$$DH_{XC,post-HFlevel}[c'_X, c'_{c,XC}|\ c_X, c_{c,XC}, c_{p,ab}, c_{p,ss}|\ \text{optional dispersion correction}]$$

Strictly speaking, for non-HF reference orbitals, a single excitations relaxation term should be added; it is omitted by Grimme and by essentially all subsequent practitioners. Van Voorhis [2008, personal communication to JMLM] assessed its importance and found its omission to have no practical consequences.

Truhlar and coworkers first coined[72] the term "doubly hybrid" for a mixture of GGA DFT correlation and pure MP2 correlation, but the term "double hybrid" as it is presently understood first was introduced by Grimme for his B2PLYP functional, which corresponded to:

**Table 1.** Error statistics (kcal/mol) for the GMTKN55 main-group benchmark and its five top-level subsets of various DFT functionals., grouped by descending rung on Jacob's Ladder. Solid horizontal lines separate between rungs, the dotted horizontal line separates between range-separated and global hybrids. Mean absolute deviation (kcal/mol) for the MOBH35 organometallic reaction benchmark is given in the last column.

| Functionals | WTMAD2 | Ther | BH | Large | Conf. | Inter | [a] MAD |
|---|---|---|---|---|---|---|---|
| ωB97M(2) | 2.19 | 0.44 | 0.26 | 0.42 | 0.58 | 0.49 | 1.9 |
| xrevDSD-PBEP86-D4 | 2.26 | 0.56 | 0.27 | 0.52 | 0.43 | 0.47 | – |
| revDSD-PBEP86-D4 | 2.33 | 0.56 | 0.31 | 0.58 | 0.41 | 0.48 | 1.5 |
| revDOD-PBEP86-D4 | 2.36 | 0.59 | 0.30 | 0.59 | 0.41 | 0.47 | 1.4 |
| revDSD-PBEP86-D3 | 2.42 | 0.54 | 0.31 | 0.55 | 0.46 | 0.57 | 1.7 |
| revDSD-PBEP86-NL[45] | 2.44 | 0.55 | 0.30 | 0.55 | 0.47 | 0.57 | – |
| revDSD-PBE-D4 | 2.46 | 0.65 | 0.35 | 0.53 | 0.43 | 0.50 | 1.7 |
| revDSD-BLYP-D3 | 2.48 | 0.57 | 0.32 | 0.57 | 0.47 | 0.55 | 2.0 |
| revDSD-BLYP-D4 | 2.59 | 0.57 | 0.34 | 0.58 | 0.48 | 0.62 | 1.8 |
| DSD-SCAN-D4 | 2.64 | 0.60 | 0.40 | 0.62 | 0.45 | 0.56 | 1.6 |
| DSD-PBE-D4 | 2.64 | 0.61 | 0.39 | 0.56 | 0.53 | 0.54 | 2.4[c] |
| DSD-PBEP86-NL | 2.64 | 0.58 | 0.40 | 0.57 | 0.54 | 0.56 | 2.1 |
| DSD-PBEP86-D4 | 2.65 | 0.54 | 0.37 | 0.63 | 0.55 | 0.56 | 1.9 |
| revDSD-PBEB95-D4 | 2.70 | 0.64 | 0.31 | 0.45 | 0.78 | 0.52 | 1.6 |
| revωB97X-2[45] | 2.80 | 0.58 | 0.37 | 0.58 | 0.50 | 0.77 | 2.3[b] |
| DSD-BLYP-D4 | 2.83 | 0.58 | 0.38 | 0.59 | 0.68 | 0.60 | 3.3 |
| B2NC-PLYP-D3 | 2.96 | 0.63 | 0.50 | 0.61 | 0.62 | 0.60 | 1.8 |
| noDispSD-SCAN$_{69}$[45] | 2.98 | 0.58 | 0.52 | 0.67 | 0.48 | 0.73 | 2.6 |
| ωB97X-2(TQ) | 2.98 | 0.59 | 0.36 | 0.59 | 0.50 | 0.93 | 2.4[b] |
| DSD-PBEP86-D3 | 3.10 | 0.55 | 0.45 | 0.49 | 0.65 | 0.97 | 2.2 |
| B2GP-PLYP-D2 | 3.14 | 0.65 | 0.43 | 0.57 | 0.80 | 0.70 | – |
| DSD-PBE-D3 | 3.17 | 0.66 | 0.41 | 0.54 | 0.73 | 0.83 | 2.4 |
| B2GP-PLYP-D3 | 3.19 | 0.63 | 0.42 | 0.66 | 0.64 | 0.85 | 2.3 |
| DSD-BLYP-D3 | 3.21 | 0.61 | 0.34 | 0.74 | 0.69 | 0.82 | 2.6 |
| SOS0-PBE0-2-D3 | 3.46 | 0.75 | 0.55 | 0.85 | 0.67 | 0.65 | 1.7 |
| B2NC-PLYP(noD) | 3.56 | 0.65 | 0.48 | 0.69 | 0.76 | 0.99 | 1.7 |
| SCS-dRPA75-D3BJ | 3.58 | 1.03 | 0.38 | 0.51 | 0.87 | 0.80 | – |
| B2PLYP-D3 | 3.90 | 0.79 | 0.63 | 0.96 | 0.74 | 0.78 | 3.0 |
| SOS1-PBE-QIDH-D3[46] | 3.84 | 0.89 | 0.45 | 0.87 | 0.77 | 0.87 | 1.9 |
| SCAN0-2[47] | 4.69 | 0.98 | 0.66 | 1.08 | 0.95 | 1.03 | 2.9 |
| ωB97M-V | 3.29 | 0.73 | 0.45 | 0.64 | 0.90 | 0.57 | 1.7 |
| ωB97M-D3[48] | 3.76 | 0.74 | 0.41 | 0.82 | 0.89 | 0.90 | 1.9 |
| ωB97X-V | 3.96 | 1.02 | 0.56 | 1.07 | 0.73 | 0.58 | 2.0 |
| ωB97X-D3[48] | 4.39 | 1.08 | 0.49 | 0.92 | 0.88 | 1.01 | 2.3 |
| CAM-B3LYP-D3[31] | 5.32 | 1.13 | 0.88 | 1.26 | 1.24 | 0.81 | 2.4[c] |
| LC-ωPBEh-D3[49] | 5.49 | 1.32 | 0.95 | 1.24 | 1.13 | 0.84 | – |
| revM11[50] | 5.73 | 1.12 | 0.76 | 1.28 | 1.61 | 0.95 | 2.7[b] |
| M11[51] | 6.42 | 0.96 | 0.57 | 1.10 | 2.54 | 1.25 | 2.7 |
| CAM-QTP00-D3[52,53] | 6.48 | 1.65 | 1.08 | 1.28 | 1.28 | 1.20 | 4.5 |
| CAM-QTP01-D3[52,53] | 6.81 | 1.26 | 0.94 | 1.21 | 1.81 | 1.60 | 2.6 |
| CAM-QTP02-D3[53,54] | 7.30 | 1.32 | 0.92 | 1.27 | 1.91 | 1.88 | 3.1 |
| M06-2X[29] | 4.79 | 0.86 | 0.48 | 1.08 | 1.22 | 1.14 | 3.1 |
| revM06[53] | 5.30 | 1.01 | 0.52 | 1.15 | 1.67 | 0.94 | 2.3[b] |
| revPBE0-D3[27] | 5.43 | 1.37 | 0.96 | 1.06 | 1.13 | 0.91 | 2.8 |
| PW6B95-D3[55] | 5.49 | 1.05 | 0.79 | 1.49 | 1.31 | 0.86 | 2.4 |
| BHandHLYP-D3[25] | 5.54 | 1.58 | 0.82 | 1.26 | 0.96 | 0.92 | 3.8 |
| MN15-D3[56] | 5.77 | 1.00 | 0.52 | 1.02 | 2.13 | 1.11 | 2.5 |
| SCAN0-D3 | 6.23 | 1.66 | 1.09 | 1.17 | 1.05 | 1.25 | 2.3 |
| B3LYP-D3 | 6.50 | 1.31 | 1.14 | 1.66 | 1.15 | 1.24 | 3.8 |
| PBE0-D3 | 6.55 | 1.38 | 1.21 | 1.37 | 1.26 | 1.34 | 2.6 |
| SCAN0[47] | 7.69 | 1.64 | 1.03 | 1.32 | 1.71 | 1.99 | 2.3 |
| M06-D3(0)[29] | 7.75 | 1.15 | 0.64 | 1.46 | 2.98 | 1.53 | 3.7 |
| B97M-V[57] | 6.37 | 1.20 | 1.00 | 1.56 | 1.76 | 0.85 | 2.9 |
| SCAN-D3[23] | 7.95 | 1.67 | 1.95 | 1.30 | 1.30 | 1.73 | 3.8 |
| revTPSS-D3[58] | 8.42 | 1.94 | 2.04 | 1.77 | 1.33 | 1.35 | 4.4 |
| TPSS-D3[22] | 9.14 | 1.84 | 2.14 | 2.02 | 1.61 | 1.53 | 4.4 |
| revPBE-D3[59] | 8.34 | 1.77 | 2.04 | 1.71 | 1.55 | 1.27 | 5.0 |
| B97-D3[60] | 8.61 | 1.82 | 1.73 | 2.28 | 1.54 | 1.24 | 5.4 |
| rPBE-D3[61] | 10.42 | 2.06 | 2.30 | 1.61 | 1.57 | 2.89 | – |
| PBE-D3[62] | 10.44 | 2.09 | 2.41 | 2.01 | 1.88 | 2.05 | 4.7 |
| PBEsol-D3[63] | 14.28 | 2.91 | 3.11 | 2.25 | 3.12 | 2.88 | – |
| SPW92[13,16] | 22.67 | 4.46 | 4.05 | 2.97 | 5.09 | 6.12 | 7.1[c] |

D3 is shorthand for D3(BJ) throughout this table. Parameters taken from the ESI of the GMTKN55 paper[38] unless indicated otherwise or defined in original paper. For the avoidance of doubt, all data in the table were taken either from Refs. [45, 53] or calculated in this work using Q-CHEM 5.2,[64] unless indicated otherwise. [a] Iron & Janes revised MOBH35 transition metal barriers[65] [b] Recalculated in present work [c] Ref. [53]

$$\mathrm{B2PLYP}: \ \mathrm{DH}_{\mathrm{BLYP,PT2}}[c_{\mathrm{HF}},\ 1-c_2 |\ c_{\mathrm{HF}},\ 1-c_2,\ c_2,\ c_2|\mathrm{null}]$$

where he obtained $c_{HF}=0.53$, $c_2=0.27$ through minimization of the error in the G2-1 atomization energies with the def2-QZVP basis set.

Later, he realized that B2PLYP captured part, but not all, of the dispersion energy in noncovalent interactions, and added the D2 dispersion correction[73] to obtain[74]

$$\mathrm{B2PLYPD}: \ \mathrm{DH}_{\mathrm{BLYP,PT2}}[c_{\mathrm{HF}},\ 1-c_2 |\ c_{\mathrm{HF}},\ 1-c_2,\ c_2,\ c_2|\mathrm{D2}(s_6=0.55)]$$

Compared to B3LYP-D3, B2PLYP-D3 cuts WTMAD2 almost in half. Moreover, with just two adjustable parameters in the electronic structure part, it handily outperforms the best global hybrid in Table 1, M06-2X[29] with 29 empirical parameters.

Yet during our own numerical explorations at the time, we found B2PLYP(D) had room for improvement for barrier heights (a problem in DFT that had preoccupied us for some time[30]). In addition, we realized that double hybrids inherit some of the slow basis set convergence of correlated WFT methods, and hence the incomplete basis set (plus problematic accuracy of the G2 reference data, as commented on repeatedly[75]) introduces some parametrization bias.

Hence we set about determining new DBH24 and W4-08 reference data for barrier heights and atomization energies, respectively, through the W4 high-accuracy WFT protocol.[76,77] From two-dimensional contour plots in $c_{HF}$ and $c_2$ of both datasets, we learned that (a) RMSD for atomization energies has not so much a minimum as an "optimum canal" spanning roughly from (0.50,0.22) to (0.72,0.42); (b) RMSD for barrier heights has a more clearly defined minimum (0.64,0.30) at the bottom of an elliptical valley; (c) (0.65,0.36) is a good compromise between these two.

The scaling parameter $s_6=0.40$ for the D2 disperison correction was subsequently determined from the original[78] S22 noncovalent interactions dataset in the same manner as Ref. [74]. Thus we finally obtained the B2GP-PLYP-D2 (GP for "general purpose") double hybrid[79] with $c_{HF}=0.65$, $c_2=0.36$, and $s_6=0.40$. As can be seen in Table 1, this represents a significant improvement over B2PLYP not just for barrier heights but also for basic thermochemistry and large-molecule reactions. It still outperforms the best rung 4 functional to date, ωB97M-V. Intriguingly, B2GP-PLYP is fairly insensitive to the type of dispersion correction it is paired with.

Six years later, Yu[80] proposed B2NC-PLYP with $c_{HF}=0.81$, $c_2=0.55$, and no dispersion correction, claiming it would be unneeded because of the high percentage of MP2 correlation. With D3(BJ) parameters from the ESI of the GMTKN55 paper,[38] however, WTMAD2 drops from 3.56 to 2.96 kcal/mol, the latter being the lowest value for any simple double hybrid.[48]

To what degree do double hybrids (particularly DSD) offer added value over simple MP2 or spin-component-scaled MP2?[81] (For a review of spin-component-scaled perturbation theory, see Ref. [82]) This issue was previously addressed in Ref. [40] for GMTKN30 and in Mehta *et al.*[83] for GMTKN55 (see especially Tables S19 and S21 in that reference). For the sake of completeness, we supply Table 2 as an MP2 counterpart of Table 1, calculated using the same basis sets.

Even with a dispersion correction, simple MP2 is in hybrid DFT performance territory, comparable to revPBE0-D3 or PW6B95-D3. SCS-MP2 (i.e., $c_{2ab}=6/5$, $c_{2ss}=1/3$)[81] does slightly better, especially with a dispersion correction; Fink's[84] S2-MP2, (i.e., $c_{2ab}=1.15$, $c_{2ss}=0.75$, obtained[84] by maximizing the overlap between the PT1 and FCI wavefunctions for some reference systems) performs intermediately. An ad hoc minimization of WTMAD2 yields what we will term S2opt-MP2 with $c_{2ab}=1.055$, $c_{2ss}=0.623$, which most resembles S2-MP2 with the correlation scaled down a bit. This has the lowest WTMAD2 of about 4.9 kcal/mol – more than twice the value for the best double hybrid in Table 1, and still one-and-a-half times worse than the best fourth-rung functional, ωB97M-V. SOS-MP2 (spin-opposite-scaled MP2,[85] i.e., $c_{2ab}=1.3$, $c_{2ss}=0$) does far worse than the other variants in the absence of a dispersion correction, but after inclusion of a dispersion correction the gap with other methods narrows considerably (see also Section 3.3 below).

**Table 2.** Error statistics (kcal/mol) for the GMTKN55 main-group benchmark and its five top-level subsets of various DFT functionals for MP2 and various spin-component-scaled variants.

| | WTMAD2 | Ther | BH | Large | Conf. | Inter | MOBH35 def2-QZVPP | |
|---|---|---|---|---|---|---|---|---|
| | Without Dispersion | | | | | | | |
| RI-MP2 | 6.91 | 1.22 | 1.23 | 1.79 | 1.47 | 1.21 | **6.0** | |
| SCS-MP2 | 5.35 | 0.95 | 1.01 | 1.15 | 1.02 | 1.23 | **2.8**[b] | |
| SOS-MP2 | 7.77 | 1.12 | 1.28 | 1.50 | 1.60 | 2.17 | **2.5** | |
| S2-MP2 | 6.10 | 1.16 | 1.15 | 1.62 | 1.25 | 0.92 | **5.4** | |
| S2opt-MP2[a] | 4.90 | 0.94 | 1.01 | 1.18 | 0.93 | 0.84 | **3.6** | |
| | With D4 [fixed parameters: $a_1=0.4$, $a_2=3.6$] | | | | | | | |
| | WTMAD2 | Ther | BH | Large | Conf. | Inter | $s_6$ | $s_8$ |
| RI-MP2 | 5.55 | 1.17 | 1.16 | 1.50 | 0.90 | 0.82 | 0.022 | −0.511 |
| SCS-MP2 | 5.03 | 0.92 | 1.01 | 1.18 | 1.07 | 0.84 | 0.167 | −0.005 |
| SOS-MP2 | 5.74 | 1.03 | 1.15 | 1.29 | 1.33 | 0.95 | 0.323 | +0.172 |
| S2-MP2 | 5.40 | 1.13 | 1.11 | 1.42 | 0.93 | 0.81 | 0.073 | −0.434 |
| S2opt-MP2[a] | 4.86 | 0.93 | 1.01 | 1.21 | 0.96 | 0.75 | 0.139 | −0.107 |

[a]The optimum parameters for spin-component-scaled MP2 on GMTKN55 are $c_{2ab}=1.055$ and $c_{2ss}=0.623$. That is actually closest to Fink's re-derived S2-MP2 (1.15, 0.75), not to SCS-MP2. [b]Iron and Janes.[65] All remaining data calculated in present work.

Clearly, as already shown in Refs. [40,83], a double hybrid amounts to more than simply the sum of its parts.

### 3.2 DSD Double Hybrids

There is intrinsically no reason why the only semilocal XC functional that works well in a DH context should be BLYP. Yet both we ourselves and Grimme found[79,86] that, while pretty much any good *exchange* functional can be substituted for Becke88,[19] the only *correlation* functional that works well for a simple double hybrid is LYP. Now LYP is derived from the Colle-Salvetti model[87] for correlation in helium-like atoms, hence the peculiar (and unphysical) feature that the LYP correlation energy of a fully polarized uniform electron gas vanishes.[88]

In addition, Martin and coworkers found in a number of studies (e.g.,[89–91]) on noncovalent interactions that the same-spin correlation energy contains quite similar information to the dispersion term. In a detailed study[89] on the conformer surface of n-pentane, we have shown the similarity between the CCSD same-spin correlation energy surface and various empirical dispersion corrections (see Figure 2).

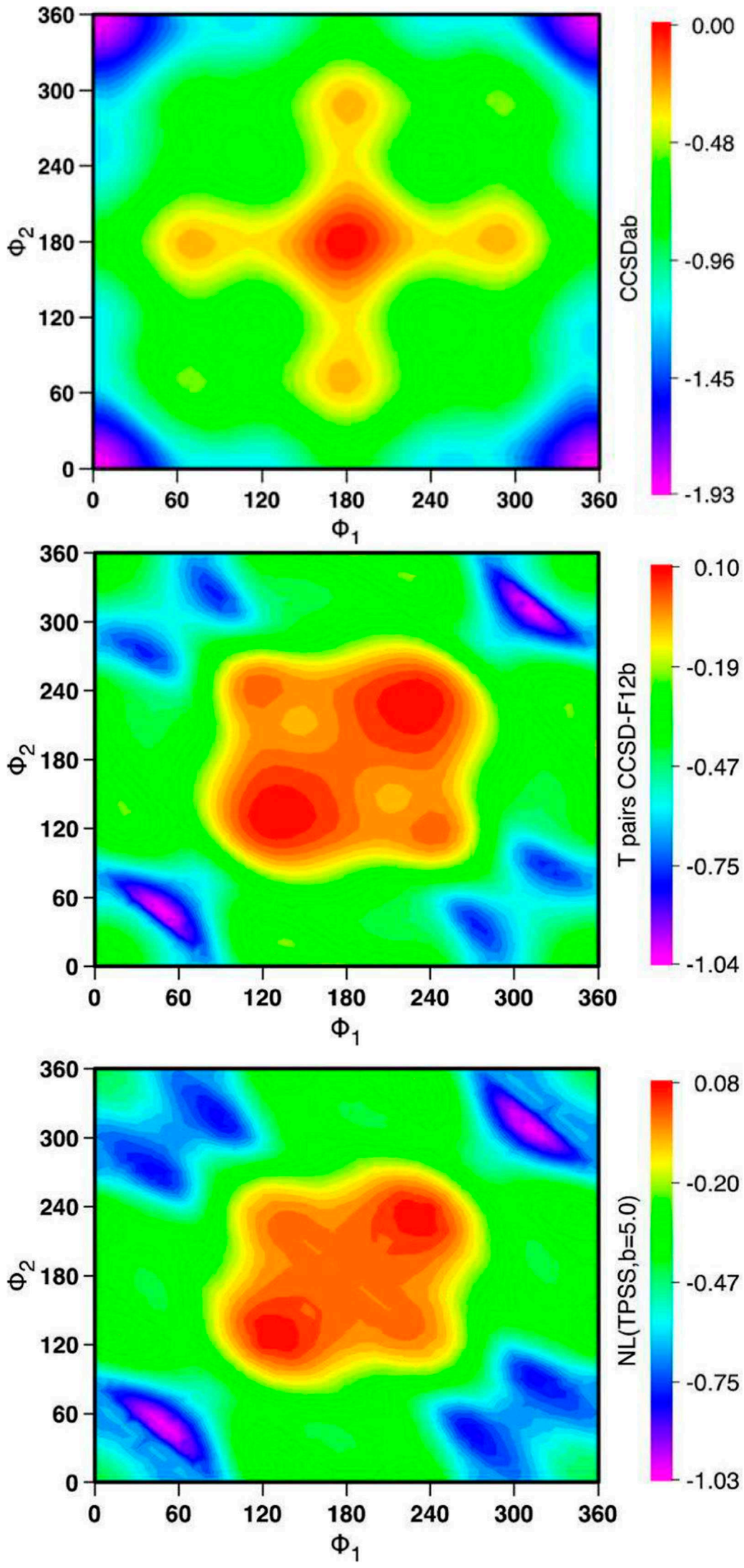

**Figure 2.** Comparison of opposite-spin correlation (top), same-spin correlation (middle), and dispersion contributions (bottom) to relative energies (kcal/mol) on the n-pentane torsion surface. From Ref. [89], courtesy of the American Chemical Society.

In combination, these factors led Kozuch and Martin[90,91] to propose the so-called DSD functional family (dispersion-corrected, spin-component-scaled, double hybrids).

$$\text{DSD-DH}:$$
$$\text{DH}_{XC,PT2}[c_X,\ c_{c,XC}|\ c_X,\ c_{c,XC},\ c_{2ab},\ c_{2ss}|D2(s_6)\ \text{or}\ D3(s_6,\ s_8=0,\ a_1=0,\ a_2)]$$

As seen in Table 1, DSD-BLYP[90] represents no major improvement over B2GP-PLYP. The major advantage of the DSD form is that it no longer requires LYP correlation specifically: Indeed, we found that even LDA correlation performs comparably to LYP, as do PBEc and PW91c, and that B95c and P86c were indeed superior to LYP. Our overall winner was the DSD-PBEP86-D3(BJ) functional, at RMSD = 1.62 kcal/mol for the training set in Ref. [91].

The explanation becomes apparent on inspection of the optimized $c_{2ab}$ and $c_{2ss}$, or rather their ratio $c_{2ss}/c_{2ab}$. For DSD-LDA, we have 0.11/0.58, for DSD-BP86 0.24/0.49, for DSD-PBE 0.12/0.53, but for DSD-BLYP 0.43/0.46. In other words, the only case where the constraint $c_{2ss}=c_{2ab}$ of simple DHs does not amount to a "Procrustean bed" is DSD-BLYP.

In refs. [91,92], using a relatively small and narrow training set, the two "winners" were DSD-PBEB95-D3(BJ) followed by DSD-PBEP86-D3(BJ). However, we see from Table 1 that on a broader sample, DSD-PBEP86-D3(BJ) clearly acquits itself better. Over the GMTKN55 set, DSD-PBEP86-D3(BJ), the original DSD-BLYP-D3(BJ), B2GP-PLYP-D3(BJ), and B2NC-PLYP-D3 consistently show up as the best performers, with WTMAD2 (weighted mean absolute deviations) in the 3 kcal/mol range.

Practical implementations of DSD-PBEP86, alas, yield slightly different total energies between different codes. The reason for that lies in the P86 correlation functional, where all codes do implement the same GGA correction but different codes apply it to different LDA correlation parametrizations. For instance, Gaussian by default uses the Perdew-Zunger[15] 1981 (PZ81) parametrization (as Perdew himself did in the original P86 paper[20]), but this has not been implemented in ORCA,[93] which instead offers either the Vosko-Wilk-Nusair[14] parametrizations (VWN3 or, preferably, VWN5) or the Perdew-Wang[16] 1992 (PW92) parametrization. DSD-PBEPBE does not have the same reproducibility issue, since all codes known to the author use PW92 local correlation with the PBEc correction.

Do mGGAs offer any advantage over GGAs for the semilocal part of a DH? The comparatively good performance of DSD-LDA suggests otherwise, and indeed Kozuch and Martin[91,92] found that DSD-TPSS offers no advantage over DSD-PBE or other DSD-GGAs. The only mGGA correlation that looked promising was B95c,[94] in which the only "meta" aspect is a factor in the same-spin correlation term that ensures one-electron systems have no spurious self-correlation.

Roch and Baldridge introduced the mSD family[95,96] of functionals (without dispersion correction) featuring just two independent parameters: the percentage of HF exchange and the percentage of SCS-MP2 correlation, while they constrain the ratio $c_{2ab}:c_{2ss}=6/5:1/3$ as in the original SCS-MP2 paper.[81] Aside from different ratios (e.g., 1.15 : 0.75 as in S2-MP2[84]) being justifiable on theoretical grounds[82] *even for HF reference orbitals*, there is no intrinsic reason why either the SCS-MP2 or S2-MP2 ratios would necessarily be valid or optimal for GLPT2 in a basis of Kohn-Sham orbitals.

As shown in Figure 2 of Ref. [83], mSD-PBEPBE performance for GMTKN55 is not just inferior to other double hybrids, but it is indeed comparable to an mGGA.

### 3.3 Dispersion Corrections in a DH Context

Among dispersion corrections that have been used in conjunction with double hybrids are the original D2,[73] D3 with zero-damping,[60] D3 with Becke-Johnson damping,[97] D4,[98,99] and the Vydrov-Van Voorhis[100] nonlocal dispersion functional. A detailed discussion of different empirical dispersion approaches is beyond the scope of this review: the reader is referred to, e.g., Refs. [101–103]. It suffices to say in this context that the original D2 was a damped Lennard-Jones correction with a single parameter (a prefactor $s_6$) and fixed atomic parameters that were completely insensitive to the chemical environment. In D3, a bond order dependence is introduced, while D4 is even more responsive to the chemical environment through a dependence on partial charges. Additionally, D4 includes a 3-body Axilrod-Teller-Muto[104,105] type term.

While Refs. [91,92] do suggest that double hybrids are more forgiving of the foibles of the simple D2 method than DFT functionals on lower rungs would be, D3(BJ) does generally lead to better results. Simply using D4 as a drop-in replacement for D3BJ improves results in many cases (e.g., DSD-BLYP-D4, DSD-PBEP86-D4, and especially DSD-PBE-D4, Table 1) while the reoptimized revDSD-XC-D4 functionals almost invariably outperform their revDSD-XC-D3(BJ) cognates.[45]

Conversely, one might decide to eliminate the dispersion correction altogether and instead rely on same-spin correlation for the same purpose. In the so-called DSD-XC-noD[92] or noDispSD-XC[92] functionals, exactly this choice is made: typically, $c_{2ss}$ there is about 0.2–0.3 larger than in the corresponding DSD-XC-D3(BJ) functional.

Among global double hybrids without dispersion correction, noDispSD-SCAN is the best performer. If range-separated exchange is allowed, however, a partial reparametrization[45] of ωB97X-2,[106] which we denote revωB97X-2,[45] is the overall winner at WTMAD2 = 2.80 kcal/mol (Table 1).

We also optimized $s_6$ and $s_8$ coefficients (prefactors for the $R^{-6}$ and $R^{-8}$ terms, respectively) in the D4 dispersion correction[99] to go with spin-resolved MP2 variants (Table 3, lower half). Dispersion is absolutely necessary for SOS-MP2, while for simple MP2 and S2-MP2, we find $s_6$ coefficients for the $R^{-6}$ term nearly zero combined with significant *negative* coefficients for the $R^{-8}$ term. This indicates compensation for an overcorrection at medium-range in MP2, as previously noted in Ref. [107]. A similar pattern – small $s_6$ and large negative $s_8$ – can be seen in Table S1 of Mehta *et al.*[83] for MP2-D3(BJ), as well as in Table 14 of Ref. [107] (a revision of the S66x8 noncovalent interactions benchmark[108]). The latter reflects that full MP2, in a noncovalent interactions setting, behaves correctly at long distance but overbinds at intermediate distance.

### 3.4 Reparametrizing for Chemical Robustness: revDSD

For reasons of computational efficiency, the original training set[91,92] for DSD-PBEP86 *et al.* had to be quite small: it consisted of just the W4-08 total atomization energies,[79] the DBH24 barrier heights,[109] the S22 noncovalent interactions set,[110,111] Truhlar's model system[112] for the Grubbs metathesis, our older set[113] of prototype oxidative additions at Pd atom, and Korth and Grimme's "mindless benchmark.[114]"

Experimentation with weights for these subsets, as well as with the addition of a rare-gas dimers benchmark, revealed that sensitivity to the weights assigned to the Pd, S22, and RG subsets was greater than desirable. This prompted the question whether reparametrization to a very large and chemically diverse dataset like GMTKN55[38] (see Section 2 above) would yield more robust parametrizations.[1]

Many of the parameters in the double hybrid, and hence in the objective function, are entirely or nearly linear in the optimization sense of the word. Hence, we carried out macroiterations on the few nonlinear parameters we do have (primarily the percentage of HF exchange) and at each point evaluated all 2,459 energies broken down into components, then employed the latter in microiteration cycles where the remaining parameters were self-consistently optimized. As the cost of re-evaluating all dispersion corrections is negligible compared to the CPU time for the electronic structure calculations, we attempted to also microiterate the nonlinear parameters in the dispersion correction. We however found that this leads to a numerically highly unstable optimization as the penalty surface in those parameters is quite flat. We therefore instead use fixed, reasonable values.

Across the board, we find our revised DSD functionals (denoted revDSD) to have lower WTMAD2 values than the originals, but the difference is particularly blatant for revDSD-PBEP86-D4. With just six adjustable parameters, said functional approaches the performance of ωB97M(2) to within overlapping uncertainties.

One notable change between original and revised optimized parameters is that same-spin coefficients drop substantially: this is advantageous if one desires to retain only the opposite-spin MP2-like term for reasons of computational efficiency (see below).

Finally, we had already found earlier (see Figure 1 in Ref. [92]) that performance of a double hybrid for the W4-11 thermochemical benchmark[44] depends much more weakly on the fraction of HF-like exchange than an ordinary hybrid, and that the dependence is weaker still for DSD double hybrids. In the revDSD paper we showed that this is true for GMTKN55 as well if all remaining parameters are reoptimized self-consistently: for DSD-SCANx-D3, we see that varying $c_X$, the fraction of HF-like exchange, over the range 0.63 to 0.74 affects WTMAD2 by less than 0.05 kcal/mol. Hence, semi-arbitrary values can be chosen such as the published $c_X$ for the original DSD functional – or, failing that, $0.69 \approx 3^{<-1/3}$ as a reasonable compromise value.

[1] It is worth mentioning that the unmodified WTMAD2 diagnostic displays some hypersensitivity to computational details in the RG18 (rare gas dimers and oligomers) and HEAVY28 (bond separation reactions of heavy p-block elements) subsets, owing to the small reaction energies involved.

### 3.5 Accelerating Double Hybrids and Elimination of Same-Spin Correlation

An objection often raised against double hybrids is their alleged high computational cost. However, this is only really an issue with codes that do not allow DF-MP2, a.k.a. RI-MP2[115,116] (density fitting MP2, "resolution of the identity" MP2).

RI-MP2 still asymptotically has an $O(N^5)$ scaling with system size. However, in our calculations on the GMTKN55 set, we found that the RI-MP2 step reached at most 25–30 % of total CPU time (for $C_{60}$ isomers and larger). For small molecules, the RI-MP2 step's cost is a single-digit percentage of the total. Still, for large molecules, scaling reduction would be highly desirable.

There are a number of techniques involving localized pair natural orbitals (such as DLPNO-MP2 of Neese, Valeev, and coworkers[117,118]) that asymptotically scale as O(N) for an approximate correlation energy.

If one eschews orbital localization, Head-Gordon and coworkers[85] showed that the Häser-Almlöf[119] Laplace transform MP2 technique can be used to evaluate just the opposite-spin MP2 term (SOS-MP2) in $O(N^4)$ time. Variations on this approach have been implemented in several electronic structure codes.[120–123] Very recently, the tensor hypercontraction approach of Song and Martinez[124,125] offers an $O(N^3)$ alternative.

(Other options are atomic orbitals MP2 with distance screening as advocated by Ochsenfeld and coworkers[126] or stochastic MP2.[127])

The first attempted double hybrid with $c_{2ss}=0$ was B2-OS3LYP by Head-Gordon and coworkers,[128] followed by PWPB95 and PTPSS from the Grimme group.[40]

In Ref. [92] we provided alternative fits that are same-spin-free, which we denoted DOD-PBE-D3, DOD-PBEP86-D3, and the like.

$$\text{DOD-DH}: \\ \text{DH}_{\text{XC,PT2}}[\text{c}_\text{X},\ \text{c}_{\text{c,XC}} |\ \text{c}_\text{X},\ \text{c}_{\text{c,XC}},\ \text{c}_{2\text{ab}},\ 0 | \text{D2}(\text{s}_6) \text{ or} \\ \text{D3}(\text{s}_6,\ \text{s}_8 = 0,\ \text{a}_1 = 0,\ \text{a}_2)]$$

It stands to reason that degradation in performance compared to the corresponding DSD functionals would be smallest for semilocal functional components associated with a small optimum $c_{2ss}$, such as DSD-PBE-D3. In our recent study Ref. [45], however, we found that DSD-SCAN-D3(BJ) and the refitted revDSD-PBEP86-D3(BJ) both had small $c_{2ss}$, and that their analogues DOD-SCAN-D3(BJ) and revDOD-PBEP86-D3(BJ) yield essentially the same performance without same-spin MP2 correlation.

### 3.6 XYG3 Type Functionals and xDSD

Zhang *et al.*[129] argued that the damped semilocal correlation used during the orbital step will adversely affect the orbitals, both in terms of shape and in terms of orbital energies. They therefore proposed the XYG3 functional, which generally corresponds to:

$$\text{XYG3}: \\ \text{DH}_{\text{B3LYP,PT2}}[\text{B3LYP}|0.8033,\ 0.2107,\ 0.3211,\ 0.3211|\text{null}]$$

(In fact, the exchange part is $0.8033E_{X,HF} + 0.1967E_{X,Slater} + 0.2107(E_{X,B88}\text{-}E_{X,Slater})$.)

An SOS version[130] has the form

$$\text{XYG3}-\text{OS}: \\ \text{DH}_{\text{B3LYP,PT2}}[\text{B3LYP}|0.7731,\ 0.2309\text{VWN3} + 0.2754\text{LYP}, \\ 0.4364,\ 0|\text{null}]$$

Further additions to this family have been made and reviewed in detail in Ref. [131].

In Section 5.3 of Ref. [40], Goerigk and Grimme (GG) address the effect of full vs. partial semilocal correlation in the orbital generation step, as well as of using low vs. high percentages of HF exchange in that step. They show that in fact the orbitals used in B2PLYP are not qualitatively different from BHalf&HalfLYP orbitals, and that the primary effect of reducing the fraction of HF exchange is not the shapes of the occupied orbitals (which are relatively invariant across XC functionals[132]) but an artificial lowering of the virtual orbital energies. As a result, MP2 denominators are smaller and the MP2 energy is artificially raised. GG attribute the relatively good performance for main-group chemistry of XYG3 and friends to the fact that orbitals generated with low percentages of HF exchange are less prone to spin contamination.[2]

In a later collaboration, a "nonempirical" xDH-PBE0 was proposed.[133] In a comment[134] on said paper, we explored an XYG3-like modification of our *empirical* DSD functionals, e.g.:

$$\text{xDSD}: \\ \text{DH}_{\text{XC,PT2}}[c_x, 1|c_x{}', \text{c}_{\text{c,XC}}\ ,\ c_{2ab}, c_{2ss}|\text{D3}(s_6, a_2, \text{s}_8 = \text{a}_1 = 0)]$$

We found that $c_x = c_x{}'$ systematically yields the best results for a given $c_x$ value, and that for a given XC (e.g.,

[2] Variational admixture of higher-spin determinants in UHF or UKS calculations, leading to lower energy solutions that are not eigenfunctions of the $S^2$ operator.

PBEP86), xDSD may be marginally more accurate than DSD. We later confirmed[45] this finding for xrevDSD-PBEP86-D3 (BJ) vs. revDSD-PBEP86-D3(BJ): the WTMAD2 lowering of about 0.1 kcal/mol appears to be principally due to the RSE43 radical reaction subsets.

In a later article comparing xDH-PBE0 and DSD-PBEPBE-D3(BJ) for different properties, it was argued[135] that xDH-PBE0 had an edge in terms of charge delocalization, and ascribed this to allegedly reduced self-interaction error (SIE).[3] One way to thoroughly reduce SIE would be the use of range-separated hybrids, which we will discuss presently.

### 3.7 Range-Separated Double Hybrids and ωB97M(2)

In their most commonly used form, the interelectronic repulsion $r_{12}^{-1}$ is split up by means of an error function of $r_{12}$ into a short-range(SR) and a long-range(LR) component

$$\frac{1}{r_{12}} = \underbrace{\frac{1-[\alpha+\beta\mathrm{erf}(\omega r_{12})]}{r_{12}}}_{\text{SR= Short Range}} + \underbrace{\frac{\alpha+\beta\mathrm{erf}(\omega r_{12})}{r_{12}}}_{\text{LR= Long Range}}$$

$$\text{where, } 0 \leq \alpha+\beta \leq 1$$

$$0 \leq \alpha \leq 1$$

$$0 \leq \beta \leq 1$$

One of the earliest such functionals to gain wide acceptance was CAM-B3LYP of Handy and coworkers,[31] where $\alpha=0.19$; $\beta=0.46$; $\omega=0.33$.

Another common such functional is LC- ωPBE where $\alpha=0$ and $\beta=1.0$

There are several strategies for setting the range-separation parameter ω. One is to tune it for every system[136] to minimize the difference between the Koopmans' Theorem and ΔSCF values of the IP and the EA, thus ensuring the HOMO-LUMO gap approaches the "fundamental gap" IP-EA. While this strategy is very valuable for optical spectroscopy or materials properties, its application to thermochemistry (or, generally, any "mass production" computational project) is intrinsically awkward. For the CAM-QTP functionals,[52,137] Bartlett and coworkers instead carried out the tuning for a small set of reference molecules, then assumed these parameters to be transferable to all systems.

In an attempt to walk a middle course between the Perdew school of nonempirical functionals and the heavily parametrized functionals of the Truhlar group (such as M06,[138] M11,[51] and MN15[56]), Mardirossian and Head-Gordon developed a combinatorial optimization procedure in which they do expand exchange and correlation functionals into Becke-Handy[28,139] power series, but use a very large training set and winnow terms in the series expansions down for statistical significance. In this manner, they climbed up the Jacob's Ladder with the mGGA B97M-V,[57] then the range-separated hybrid GGA ωB97X-V[32] ($\alpha=0.167$, $\beta=1.0$, $\omega=0.3$, WTMAD2 = 3.96 kcal/mol) and the RSH meta-GGA ωB97M-V.[33] The latter was found ($\alpha=0.15$, $\beta=1.0$, $\omega=0.3$, WTMAD2 = 3.29 kcal/mol) by Goerigk *et al.*[48] and by ourselves[45] to be the most accurate 4th-rung functional to date. ωB97X-V and ωB97M-V have 10 and 15 linear adjustable parameters, respectively – about one-third as many as in the M11 RSH mGGA functional of Truhlar and coworkers,[51] its 2019 revision,[50] or the MN15 empirical global hybrid mGGA.

With the recent ωB97M(2) functional,[140] the same "combinatorial optimization" machinery that led[32,33,57] to B97M, ωB97X-V, and ωB97M-V was applied to obtain a double hybrid. In Ref. [45] we found it to have the lowest WTMAD2 thus far reported,[45] 2.18 kcal/mol. However, the WTMAD2 differences with revDSD-PBEP86-D4 and especially xrevDSD-PBEP86-D4 are within the uncertainty of the reference data, while these latter functionals only have 6 adjustable parameters (just 5 in the DOD variants) rather than the 16 in ωB97M(2).

In an attempt to see if we could reduce WTMAD2 further by using range-separated exchange, we found a preliminary ωDSD-PBEP86-D3(BJ) with $\omega=0.16$; $c_{x,HF}=0.69$; $c_{c,DFT}=0.367$; $c_{2ab}=0.649$; $c_{2ss}=0.143$; $s_6=0.389$; $s_8=a_1=0$; $a_2=5.5$. Its WTMAD2 = 2.16 kcal/mol in fact slightly surpasses ωB97M(2); further research in this direction is in progress.

### 3.8 Comparison with WFT-Based Composite Ab Initio Methods

The mind wonders how WTMAD2 values of just above 2 kcal/mol compare with what can be achieved using composite wavefunction thermochemistry protocols (see, e.g., Refs. [141,142] for recent reviews), such as the Gn family,[143,144] CBS-QB3,[145,146] or the ccCA approach.[147] While evaluating the entire GMTKN55 dataset with them was deemed computationally too costly, we have calculated WTMAD2 for a subset of small and medium-sized molecules that does not include elements beyond Kr (for which the basis sets are lacking). This left 642 reaction energies: we note that all geometries were frozen at the GMTKN55 reference, and that zero-point and thermal corrections were excluded. (All such calculations were performed in this work using Gaussian 16[67] on the Faculty of Chemistry's Linux cluster.)

The WTMAD2 for full G4 theory,[148] 1.76 kcal/mol, is still markedly superior to the ωB97M(2) value for this subset, 1.95 kcal/mol, while the lower-cost variants G4MP2[149] and G3B3[150] clock in at 2.29 and 2.20 kcal/mol, respectively, for the same subset; CBS-QB3 achieves 2.25 kcal/mol. In all, it is clear that the best double hybrids have now entered the accuracy regime of composite WFT methods, at lower cost and with gentler CPU time scaling. (Moreover, relatively inexpensive analytical first and second derivatives are available for the double hybrids.)

[3] Briefly, SIE refers to imperfect cancellation of diagonal term between exact Coulomb and approximate exchange.

In principle, localized pair natural orbital coupled cluster theory[7,8] may represent a linear-scaling alternative amenable to still larger molecules.[151] It should be kept in mind, however, that aside from any empirical parameters that seek to improve accuracy/"trueness," any DLPNO-CCSD(T) or PNO-LCCSD(T) implementation is reliant on several adjustable parameters that will affect numerical precision at chemically significant levels. (See, e.g., Refs. [7,152] for detailed discussions.) While one might deem such „empiricism of precision" the lesser of two evils (since one needs no reference datasets, just recalculation at tighter thresholds to establish convergence to the canonical limit), it remains a source of uncertainty nevertheless.

## 4. Basis Set Convergence and Double Hybrids

As the total energy of a double hybrid is a superposition of a hybrid DFT-like component and an MP2-like component, it follows that (a) the basis set convergence behavior would be intermediate between the geometric behavior observed for hybrid DFT and the asymptotic $L^{-3}$ behavior[153,154] (with L the highest angular momentum in the basis set) of MP2; (b) the MP2-like behavior will dominate in the large basis set limit as the hybrid DFT component will have reached saturation. (Compared to straight MP2, it is obviously mitigated by a factor $c_{2ab}$; $E_{2ss}$ converges faster, more like $L^{-5}$.[155])

It also follows that basis set dependence of double hybrids would be most WFT-like for functionals with large fractions of PT2 correlation, and (e.g., as seen in Ref. [40] for PWPB95) most DFT-like for functionals with small percentages of PT2 correlation. (We note in passing that Mardirossian and Head-Gordon[156] found Minnesota functionals to exhibit anomalously larger basis set sensitivity than other rung 4 functionals.)

For wavefunction calculations, the Dunning correlation consistent basis sets[157,158] are something of a *de facto* standard. Hence, they are efficient for the MP2-like component of a double hybrid. However, as the sp function contractions in them are based on HF orbitals, they are intrinsically less suited for pure or hybrid DFT or for the hybrid DFT component of a double hybrid.

Conversely, the pc-n and aug-pc-n polarization consistent basis sets of Jensen[159–161] are inherently well suited to DFT but perhaps less to the MP2-like component of a double hybrid.

The Karlsruhe (a.k.a. Weigend-Ahlrichs) basis sets[162] offer an interesting compromise here, as specifically the "def2" sequence was developed with both HF/DFT and correlated WFT methods in mind. Additionally, they cover the elements H–Rn (using relativistic pseudopotentials from Rb onward).

When parametrizing an empirical DFT functional, one could follow two approaches: either parametrize to the CBS (complete basis set limit) – which guarantees that results obtained with the functional will be basis set convergent – or parametrize to a basis set that is practically useful, in which case the parameters will be specific to that basis set. In DFT more generally speaking, Adamson, Gill, and Pople's EDF1 (empirical density functional one)[163] was the first to specify a particular basis set (and integration grid, SG-1). Similarly, Chai and Head-Gordon[106] propose two sets of parameters for their ωB97X-2 double hybrid: one set for extrapolation to the CBS limit and another specific to the 6-311++G(3df,3pd) "Large Pople" basis set.

Boese, Martin, and Handy[164] considered the sensitivity of parameters in empirical rung 4 functionals fitted for specific basis sets, and found the parameters to be reasonably transferable as long as the basis sets were of at least triple-zeta plus polarization quality. In terms of double hybrids, the situation is murkier, but our experience suggests that similar transferability requires a basis set of at least quadruple-zeta quality.

Radom and coworkers[165] considered basis set dependence of double-hybrid parametrizations: they concluded that, while the deficiencies of triple-zeta basis set could to some degree be 'smoothed over' by ad hoc parametrizing to the basis set, this approach no longer works for still smaller basis sets. Basis set extrapolation formulas for double hybrids have been proposed by Chuang *et al.*[166] and by Chan and Radom.[167]

Witte, Neaton, and Head-Gordon,[168] from DFT basis set convergence studies on the S22 weak interactions benchmark,[78,169] recommended to use at least a quadruple-zeta basis set for DFT parametrization. Basis set convergence for the larger S66x8 benchmark[108] has been considered in great detail in Ref. [107], particularly Tables 6–9 there (for HF, PBE0, MP2, and DSD-PBEP86-D3, respectively).[4] From both studies emerges that if one wants to avoid counterpoise[170] calculations, then def2-QZVPD will effectively represent a basis set limit for rungs 1–4; from the latter study it became clear that for DSD-PBEP86 even def2-QZVP is not adequate unless counterpoise corrections are applied. (See also Refs. [171,172] for detailed discussion of basis set superposition error for wavefunction methods.)

In the B2GP-PLYP paper, we used extrapolation from Jensen's aug-pc2 and aug-pc3 polarization consistent[160] basis sets. In the DSD papers,[91,92] we did the same for the thermochemistry subset, but for the S22 weak interactions benchmark[78,169] were forced to fall back to a rather smaller basis set. In the revDSD paper (see below), we used def2-QZVPP except for anionic subsets, where we used def-QZVPPD.

As parametrization to very large training sets becomes unwieldy for basis set extrapolation, both the Head-Gordon group[32,33,57,69] and the present authors[45] adopted the compromise solution of using a fixed def2-QZVPPD (or similar) basis set for that purpose.

[4] We note a typo in Table 6 of Ref. [107]: the half-counterpoise entry for haVQZ should read 0.021.

## 5. 'Nonempirical' Double Hybrid Functionals

There has traditionally been a major endeavor on the part of the "physicist" camp in the DFT community to develop functionals devoid of empirical parameters, purely from constraints that the exact exchange-correlation functional can be proven to obey. The SCAN (strongly constrained and appropriately normed[23]) meta-GGA of Perdew and coworkers probably has taken this approach the furthest, building on their earlier PBE[21] and TPSS[22] functionals. On rung 4, PBE0 and SCAN0 are examples.[27,173]

Hence there have been efforts to develop double hybrids without resorting to empirical parameters. By applying the adiabatic connection formalism and assuming the simplest possible forms of the integrand that has the correct curvature, Sharkas, Toulouse, and Savin[174] arrived at the single-parameter expression:

$$E_{xc} = \lambda E_{x,HF} + (1\text{-}\lambda)E_x[\rho] + E_c[\rho]\text{-}\lambda^2 E_c[\rho_{1/\lambda}] + \lambda^2 E_{(2)}$$

in which $\rho_{1/\lambda}$ denotes the coordinate-scaled density $\rho_{1/\lambda}(\mathbf{r}) = \lambda^{-3}\rho(\mathbf{r}/\lambda)$. Now three approximations can be made for $E_c[\rho_{1/\lambda}]$:

(a) $E_c[\rho_{1/\lambda}] \approx E_c[\rho]$ leads to the quadratic one-parameter double hybrid:

$$E_{xc} = \lambda E_{x,HF} + (1\text{-}\lambda)E_x[\rho] + E_c[\rho](1\text{-}\lambda^2) + \lambda^2 E_{(2)}$$

(b) Assuming that $E_c[\rho_{1/\lambda}] \approx E_{(2)}$ leads to the simple hybrid $E_{xc} = \lambda E_{x,HF} + (1\text{-}\lambda)E_x[\rho] + E_c[\rho]$

(c) Assuming instead that $E_c[\rho_{1/\lambda}] \approx \lambda E_c[\rho] + (1\text{-}\lambda)E_{(2)}$ (i.e., a linear interpolation between the "spread out"-density regime where PT2 should be a good approximation to the correlation energy, and the "compressed" density regime where the semilocal functional should be more reliable) leads to the cubic one-parameter double hybrid:[175]

$$E_{xc} = \lambda E_{x,HF} + (1\text{-}\lambda)E_x[\rho] + E_c[\rho](1\text{-}\lambda^3) + \lambda^3 E_{(2)}$$

The resulting family of "nonempirical" double hybrids has been reviewed in Ref. [176]

$$\text{PBE0}-2:\ \text{DH}_{\text{PBE,PT2}}[2^{-1/3}, 1/2|2^{-1/3},\ 1/2, 1/2, 1/2|\text{null}]$$

$$\text{PBE0}-\text{QIDH}:$$
$$\text{DH}_{\text{PBE,PT2}}[3^{-1/3}, 2/3|2^{-1/3},\ 2/3, 1/3, 1/3|\text{null}]$$

$$\text{PBE0}-\text{CIDH}:$$
$$\text{DH}_{\text{PBE,PT2}}[6^{-1/3}, 5/6|6^{-1/3},\ 5/6, 1/6, 1/6|\text{null}]$$

$$\text{PBE0}-\text{DH}:\ \text{DH}_{\text{PBE,PT2}}[1/2, 7/8|1/2,\ 7/8, 1/8, 1/8|\text{null}]$$

$$\text{SOS0}-\text{PBE0}-2:$$
$$\text{DH}_{\text{PBE,PT2}}[2^{-1/3}, 1/2|2^{-1/3},\ 1/2,\ 2/3, 0|\text{null}]$$

Alipour[177] proposed a simple way of deriving spin-opposite-scaled versions of the above, starting with the empirical [sic] observation that typically $(E_{(2)\alpha\alpha} + E_{(2)\beta\beta}) \approx E_{(2)\alpha\beta}/3$, hence $E_{(2)} \approx 4E_{(2)\alpha\beta}/3$. But in another study,[178] the same author notes that for dispersive interactions at long distance, typically $E_{(2)} \approx E_{(2)\alpha\beta}$, which actually can be justified non-empirically from SAPT analysis[179,180] of the interaction energy: at long distance, the 2nd-order exchange-dispersion term $E^{(20)}_{exch\text{-}disp}$ will vanish, leaving the spin-free $E^{(20)}_{disp}$ as the dominant term. In the present work, we have analyzed the GMTKN55 raw energy component files (raw data for the functional optimizations in our Ref. [45]) for several double hybrids; we consistently find that the ratios between unscaled $E_{(2)\alpha\alpha} + E_{(2)\beta\beta}$ and $E_{(2)\alpha\beta}$ have geometric means of the $(E_{(2)\alpha\alpha} + E_{(2)\beta\beta}) \approx E_{(2)\alpha\beta}$ ratio of about 0.30–0.32, tolerably close to Alipour's conjectured 1/3. Hence:

$$E_{xc} = \lambda E_{x,HF} + (1\text{-}\lambda)E_x[\rho] + E_c[\rho](1\text{-}\lambda^3) + (4/3)\lambda^3 E_{(2)\alpha\beta}$$

Mehta *et al.*[83] carried out a comparative assessment of empirical and nonempirical double hybrids. They found that nonempirical DHs perform considerably more poorly than their empirical counterparts, and that the only nonempirical DH they could recommend was SOS0-PBE0-2-D3(BJ), i.e., their enhancement of the SOS0-PBE0-2 double hybrid:[177] with a D3(BJ) dispersion correction:[83]

$$\text{SOS0}-\text{PBE0}-2-\text{D3(BJ)}:$$
$$\text{DH}_{\text{PBE,PT2}}[2^{-1/3}, 1/2|2^{-1/3},\ 1/2, 2/3, 0|\text{D3(BJ)}(s_6 = 0.613,$$
$$s_8 = 0.167, a_1 = 0.573, a_2 = 3.572)]$$

With our slightly modified GMTKN55 benchmark, we obtained WTMAD2 = 3.48 kcal/mol, which is actually still above the best rung 4 functional, ωB97M-V (3.30 kcal/mol).

As the famous aphorism goes, "in theory there is no difference between theory and practice; in practice, there is".[181] So for the SCAN semilocal functional[23] and the GMTKN55 dataset, we found[45] the surprisingly linear (in the $c_x = 0.50$–$0.80$ range) empirical evolution of the coefficients shown in Figure 3. (We note in passing that nonempirical SCAN-based double hybrids[47] such as SCAN0-2 have distinctly inferior performance to these DSD-SCAN empirical double hybrids, cf. Table 1)

## 6. Beyond Main-Group Energetics

### 6.1 Organometallic Reaction Barrier Heights

The GMTKN55 dataset only covers main-group systems. A large segment of transition metal systems is prone to severe static correlation, which intrinsically limits the applicability of GLPT2-based double hybrids. However, there are some benchmarks available for transition metal reactions with mild to moderate static correlation. In this section, we shall focus on the MOBH35 (metal-organic barrier heights of 35 reactions) sample of Iron and Janes,[182] particularly on its updated version[65] with basis set limit energetics from the new DLPNO-CCSD($T_1$) approach.[8] The latter, unlike the popular DLPNO-CCSD(T)[183] no longer neglects off-diagonal Fock matrix elements. The systems in MOBH35 are realistic organometallic reactions with large ligands, rather than the small model systems considered by Kozuch and Martin.[92]

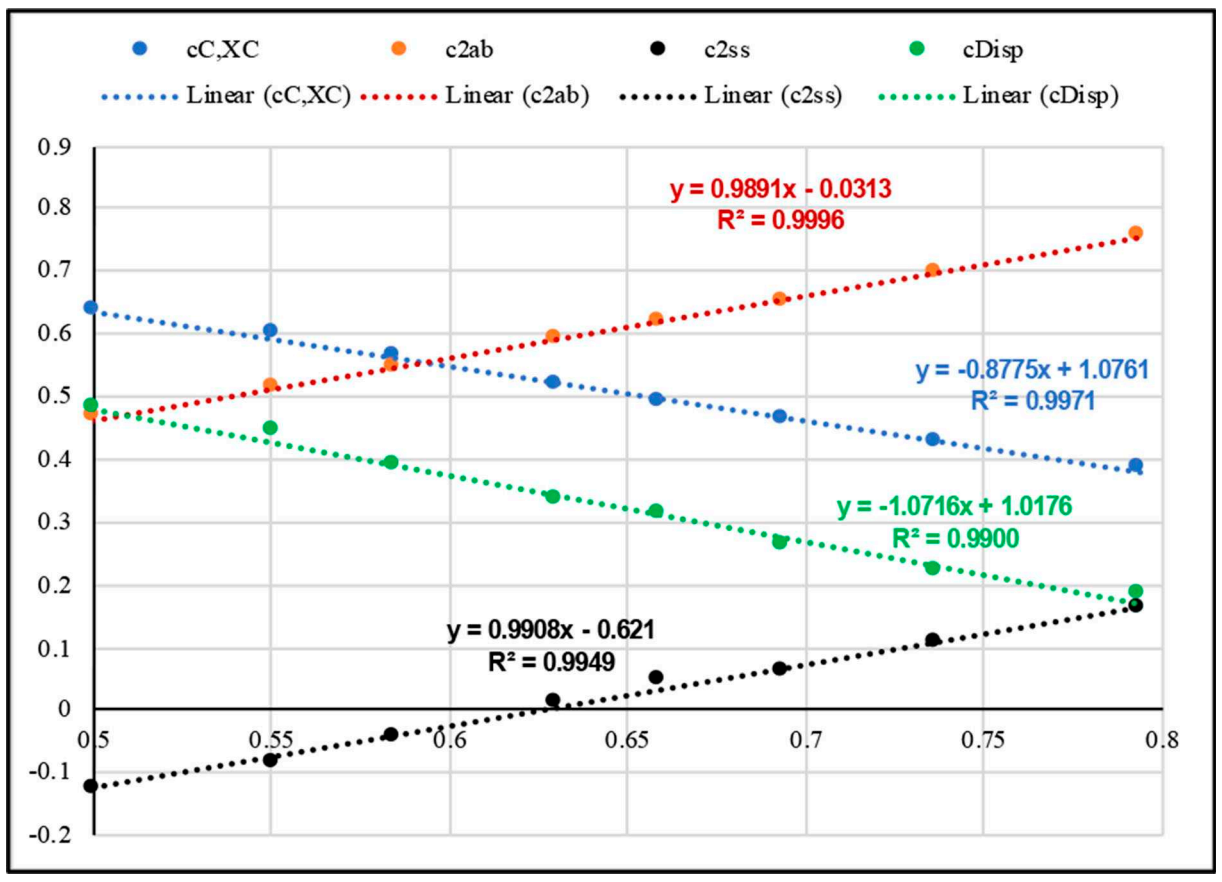

**Figure 3.** Trends in the parameters $c_{2ab}$, $c_{2ss}$, $c_{Disp}$, and $c_{C,XC}$ for the DSD-SCANx-D3(BJ) family as a function of the fraction x of HF-like exchange. From Figure S1 in ESI of Ref. [45], reused with permission, courtesy of the American Chemical Society.

As seen in the last column of Table 1, rungs 1 and 2 are clearly inadequate for this problem, while the best Rung 3 performer, B97M-V, still has an MAD = 2.9 kcal/mol. Among global hybrids, the nonempirical SCAN0[184] is the best performer, followed immediately by the empirical PW6B95-D3(BJ). RSHes seem to have a much easier time with this problem, and the best RSH for GMTKN55, ωB97M-V, also is the best MOBH35 performer on rung 4, at MAD = 1.7 kcal/mol. revDOD-PBEP86-D4 is the best rung 5 performer at 1.4 kcal/mol, a number that can be lowered to 1.2 kcal/mol by expanding the basis set to def2-QZVPP and slightly further by basis set extrapolation. We note that the revDSD reparametrizations consistently outperform the corresponding original DSD functionals for MOBH35, *despite TMs not being involved at all in parametrization.* This suggests that the revDSD functionals actually do capture the physics of the problem better than their DSD progenitors, and are not merely 'fits of round pegs to square holes'.

We also note that ωB97M(2) actually performs slightly worse than ωB97M-V, and that thus an empirical double hybrid does not automatically represent an improvement over the underlying fourth-rung functional. We would be remiss by not pointing out, however, that ωB97M-V appears to be an unusually versatile 'workhorse functional'.

This is again confirmed in a recent computational study[185] on the reaction mechanism of Milstein's[186] concurrent hydroarylation and oxidative coupling catalyzed by Ru(II) complexes. While the primary goal of this study was assessing the performance of DLPNO-CCSD(T)[8] and PNO-LCCSD(T)[7] methods compared to full canonical CCSD(T)/def2-{T,Q}ZVPP extrapolation, performance of DFT functionals was a secondary goal. The trends seen largely follow MOBH35, with again ωB97X-V and ωB97M-V putting in 'best in class' performances. revDSD-PBEP86-D3BJ still performed slightly better, however, and definitely better than the original DSD-PBEP86-D3BJ parametrization.

Some light might be shed by considering the performance of different spin-resolved MP2 variants. Somewhat intriguingly (Table 2), not only does SCS-MP2 exhibit much better performance (MAD = 2.8 kcal/mol) than standard MP2 (MAD = 6.0 kcal/mol), but SOS-MP2 without any same-spin correlation clocks in lower still at 2.5 kcal/mol. Similarly, for the transition states in Ref. [185] we find the following RMSDs from CCSD(T)/def2-TZVPP (using the same basis set): MP2 9.5, S2-MP2 8.6, SCS-MP2 5.2, and SOS-MP2 5.4 kcal/mol, the latter two figures superior to CCSD. One might be forgiven for concluding that same-spin MP2 correlation is somewhat "poisonous" in systems with significant static correlation. This finding might explain why Iron and Janes found DOD-DHs to be superior over DSD-DHs for MOBH35, and may also shed some light on the somewhat disappointing performance of ωB97M(2), for which $c_{2ab} = c_{2ss} = 0.341$.

### 6.2 Vibrational Frequencies and Derivative Properties

As analytical first and second derivatives of double hybrids are relatively inexpensive, they offer an attractive option not just for harmonic frequencies, but for anharmonic fundamentals, overtones, and combination bands obtained via second-order rovibrational perturbation theory[187] on a semidiagonal quartic force field. The latter can be obtained fairly affordably, if analytical second derivatives are available, using the Schneider-Thiel technique[188] of central numerical differentiation in normal coordinates.

In Ref. [189], performance of many exchange-correlation functionals (including double hybrids) was considered in depth for the HFREQ27 benchmark defined there. (It is quite possible, for diatomics and small polyatomics, to extract harmonic frequencies and anharmonic corrections separately from high-resolution spectra, and thus have reference values to 1 $cm^{-1}$ or better.) For basis set limit CCSD(T), the RMSD is 4.6 $cm^{-1}$, the remaining error distributed evenly between subvalence correlation (which tends to push stretching frequencies up[190]) and post-CCSD(T) valence correlation (which tends to lower them[190]).

As shown in Ref. [189], even after corrective frequency scaling MP2 has an RMSD = 30 $cm^{-1}$ near the basis set limit, compared to 32 $cm^{-1}$ for PBE0 and 25 $cm^{-1}$ for B3LYP. B2PLYP goes down to a respectable 11 $cm^{-1}$; so can B2GP-PLYP, but only after scaling by about 0.99. In contrast, DSD-PBEP86 reaches a similar accuracy without scaling, and thus its improved performance for energetic does not come at the expense of PES curvature.

Does the improved accuracy of revDSD for GMTKN55 come at the expense of harmonic frequencies? It can be seen in Figure 4 below that this is not the case, and that if anything, frequencies improve a little further – again, despite not being targeted during parametrization at all.

### 6.3 Electrical and Magnetic Derivative Properties

For NMR chemical shielding constants (CSC), it has been known since the 1990s that MP2 delivers a reasonable compromise between accuracy and computational cost.[191–194] Recently Neese *et al.*[195] implemented gauge including atomic orbital(GIAO)[196–198] CSCs for DHs in their ORCA program system, and performed a benchmark study to evaluate the performance of DHs functionals for NMR CSC calculation. Relative to CCSD(T) reference data for their chosen benchmark dataset, DSD-PBEP86-D3(BJ)[91] performed exceptionally well among all tested DHDFAs [mean absolute relative error, $MARE_\delta = 1.9\%$], thus outperforming both standard MP2 ($MARE_\delta = 4.1\%$) and SCS-MP2($MARE_\delta = 3.9\%$), as well as the commonly used meta-GGAs M06 L ($MARE_\delta = 5.4\%$) or TPSS ($MARE_\delta = 6.4\%$). In tandem with the RI-MP2 approximation, the COS (chain of spheres) approximation, and Jensen's segmented polarization consistent basis sets,[199] systems with up to ca. 400 electrons become feasible with present-day workstations: below 100 electrons, coupled cluster theory becomes a more attractive option.

In 2013, Alipour[200] considered performance of DHs for electric field response properties, specifically, isotropic and anisotropic polarizabilities of water nanoclusters. Unlike for GMTKN55-type energetics, nonempirical functionals like PBE0-DH and PBE0-2 were found to perform much better than empirical functionals: this point merits further study.

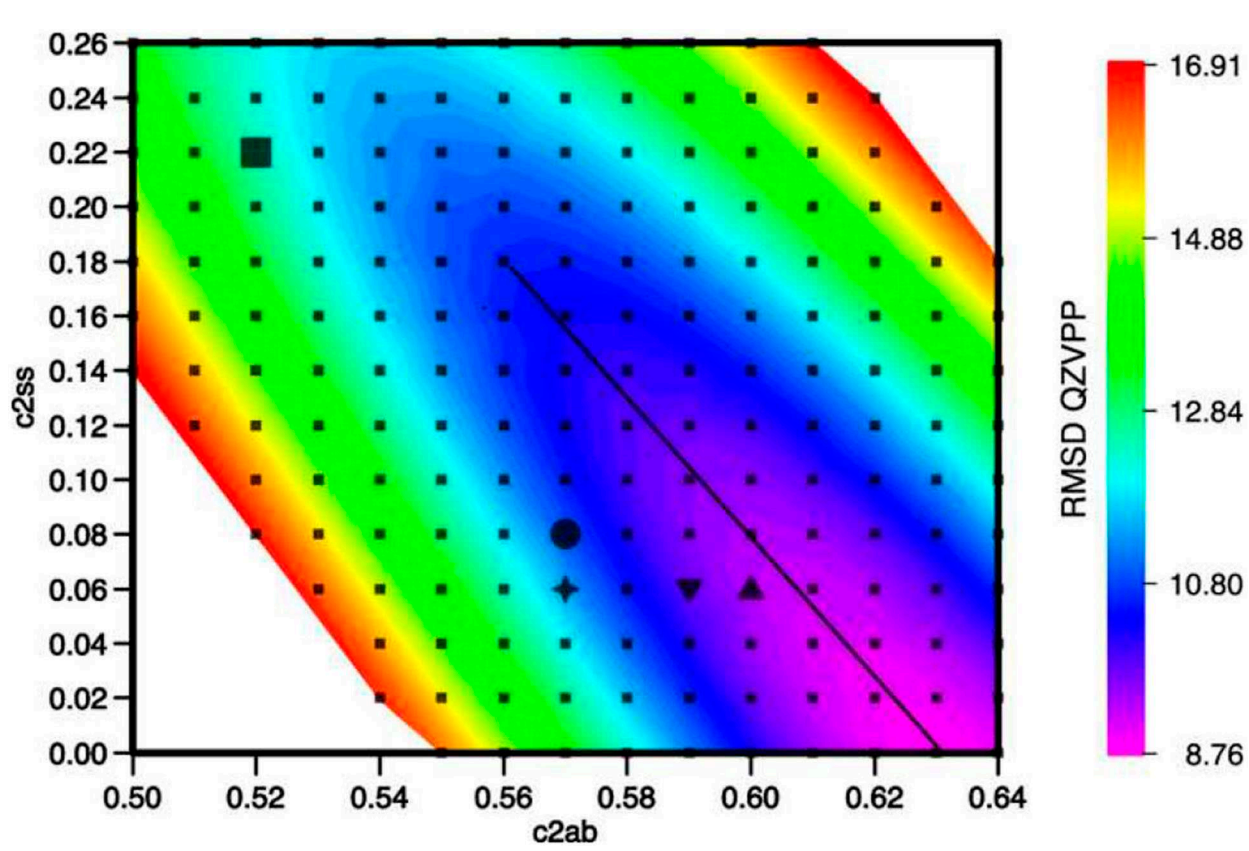

**Figure 4.** Contour plot of RMSD ($cm^{-1}$) for the HFREQ database for DSD-PBEP86-like forms, as a function of the opposite-spin and same-spin MP2 coefficients $c_{2ab}$ and $c_{2ss}$, respectively. The square marker indicates the original DSD-PBEP86-D3(BJ) solution, the large round marker revDSD-PBEP86-D3(BJ)), and the triangular ones revDSD-PBEP86-D4 on the left and revDOD-PBEP86-D4 on the right. The slanted line only serves to guide the eye. From Figure 3 of Ref. [45], reused with permission, courtesy of the American Chemical Society.

### 6.4 Electronic Excitation Spectroscopy

Already shortly after publication of the original B2PLYP paper,[201] Grimme and Neese reported[202] its extension to excited states, combining a conventional TD-DFT framework for the hybrid-GGA part with a scaled second-order perturbation correction based on Head-Gordon and Lee's CIS(D).[203] Two years later, Goerigk *et al.*[204] demonstrated promising performance of TD-B2PLYP and TD-B2GP-PLYP (both MAD = 0.22 eV) for the 142 vertical singlet excitations benchmark of Thiel and coworkers.[205] Goerigk and Grimme additionally considered low-lying valence excitations in large organic dyes[206] as well as polycyclic aromatic hydrocarbons,[207] and found particularly TD-B2GP-PLYP to perform very well, although range-separated hybrids were still superior for charge transfer states. Such global double hybrids were also found[208] to be the only ones able to simulate an exciton-coupled electronic circular dichroism (ECD) spectroscopy spectrum that had been initially studied[209] in 2008 by wave function methods.

Thus far, no spin-component scaling (SCS/SOS) of nonlocal correlation was considered in TD-DHDFAs. Schwabe *et al.* first did so in 2017.[210] Very recently, Goerigk and coworkers[211] proposed the ωB2LYP and ωB2GP-PLYP range-separated variants of B2PLYP and B2GP-PLYP,[79] respectively, with a view to application to electronic excitation energies (such as UV/VIS spectra). All parameters were frozen at their original values, with the range-separation parameter ω newly optimized to 0.30 (ωB2PLYP) and 0.27 (ωB2GP-PLYP). These authors found that particularly the latter remedies the deficiencies of the global double hybrid B2GP-PLYP for Rydberg excitations as well as charge transfer excitations, while preserving its excellent performance for valence excitations.

## 7. Orbital-Optimized Double Hybrids

Another approach that should be mentioned in passing is the adoption of orbital-optimized MP2 (OO-MP2)[212,213] inside a double hybrid framework. In such methods, the total MP2 energy is optimized with respect to the orbitals. This approach was considered first[212,213] for MP2, SCS-MP2, and SOS-MP2, then later for nonempirical double hybrids (e.g., Refs. [46,214]).

Najibi and Goerigk[215] carried out a more comprehensive survey of both empirical and nonempirical double hybrids, as well as spin-resolved MP2 variants, for a subset of GMTKN55 plus a few dedicated test sets. For the singlet-triplet splitting in the polyacene series benzene through hexacene, OO-DHs often cut errors in half compared to the underlying DH. Noticeable improvements are likewise seen for the S22 noncovalent interactions[110,111] and W4-11 small-molecule thermochemistry[44] sets, but performance for barrier heights actually gets worse. As summarized in Figure 5a of Ref. [215], for the closed-shell tests only OO-SOS-MP2 and OO-SCS-MP2 represent improvements over SOS-MP2 and SCS-MP2, and in OO-DSD-PBEP86 and OO-PBE0-2 significant *deterioration* is actually seen. However, if also open-shell cases are considered, then those two latter double hybrids actually do represent improvements, while for OO-SOS-MP2 and OO-SCS-MP2, the "OO bonus" is quite marked. Further investigation appears to be warranted.

## 8. dRPA Based Double Hybrids

The Achilles Heel of MP2-like correlation contributions are the orbital energy differences in the denominator. For molecules with small band gaps (absolute near-degeneracy, a.k.a., Type A static correlation[70]), MP2-based DHs will intrinsically struggle. Chan, Goerigk, and Radom[216] considered the use of higher-order perturbation theory and CCSD, and found that this leads to no significant improvement despite the extra computational cost.

One alternative that has been explored in the literature is RPA, the random phase approximation.[217,218] From a wave function perspective, Scuseria and coworkers[219,220] proved the equivalence of RPA with drCCD, direct ring coupled cluster with all doubles.[221]

A linear-scaling implementation of the dRPA or direct RPA approximation, in which exchange is neglected, is actually available in the MRCC program.[222,223]

Kallay and coworkers[224] then proposed the dRPA75 double hybrid:

$$\mathrm{dRPA75:}\;\; \mathrm{DH}_{\mathrm{PBE,dRPA}}[3/4, 1|3/4,\; 0, 1, 1|\mathrm{null}]$$

The original paper claimed excellent performance for noncovalent interactions, but this was due to error compensation between BSSE (basis set superposition error) with the AVTZ basis set they were using, and model incompleteness. (RPA correlation energies asymptotically converge[225] as $L^{-3}$ with the highest angular momentum $L$ in the basis set, similar to the familiar partial-wave expansion of the MP2 and CCSD correlation energy.[153,154,226,227] For practical $L=3$ and $L=4$ basis sets, Mezei *et al.*[228] found $L^{-2.5608}$, which for this basis set pair is equivalent to $(L+0.593)^{-3}$, cf. Eq. 9 in Ref. [229]) For the S66x8 noncovalent interactions benchmark,[107,108] near the basis set limit and with counterpoise corrections,[170] we found[107] that dRPA75 systematically underbinds (MSD = −0.82, RMSD = 1.09 kcal/mol).

This seems to be at odds with the received wisdom that full inclusion of RPA would make dispersion corrections redundant. All becomes clear, however, when we consider that in the same study, similar underbinding is found for CCSD/CBS (MSD = −0.59, RMSD = 0.79 kcal/mol) – and that correcting this requires either introducing (T) or adding a dispersion correction. Fitting a D2 dispersion correction dramatically improves statistics for both levels of theory: for CCSD/CBS, we found $s_6=0.225$ (RMSD = 0.16 kcal/mol), compared to dRPA75-D2, $s_6=0.314$ (RMSD = 0.15 kcal/mol). Upgrading the dispersion correction to D3(BJ) lowers RMSD for dRPA75-D3(BJ) to 0.09 kcal/mol, best-in-class.[107]

$$\begin{aligned}&\mathrm{dRPA75-D3(BJ):}\\ &\mathrm{DH}_{\mathrm{PBE,dRPA}}[3/4,\; 1|3/4,\; 0,\; 1,\; 1\\ &|\mathrm{D3(BJ)}(s_8=0.375, a_2=4.505, a_1=s_8=0)]\end{aligned}$$

Since dRPA is spin-free, spin-component scaling is pointless at least for closed-shell cases since $(E_{\mathrm{dRPA,aa}}+E_{\mathrm{dRPA,bb}})=E_{\mathrm{dRPAab}}=E_{\mathrm{dRPA}}/2$. However, for open-shell cases there will be a difference, particularly as dRPA has a spurious self-correlation energy for unpaired electrons (Section 7.2 of Ref. [230]). Taking this into account removes a serious problem with atomization energies (on account of the separated atoms).

$$\begin{aligned}&\mathrm{SCS-dRPA75:}\\ &\mathrm{DH}_{\mathrm{PBE,dRPA}}[3/4, 1|3/4,\; 0, a_{ab},\; 2-a_{ab}|\\ &\mathrm{D3(BJ)}(s_8=0.375, a_2=4.505, a_1=s_8=0)]\end{aligned}$$

where the opposite-spin scaling factor $a_{ab}=1.26$ at the CBS limit and 1.50 for the AVTZ basis set, and setting the same-spin scaling factor to 2-$a_{ab}$ ensures that the method reduces to dRPA75 for closed-shell cases.

In the present work, using the implementation in the MRCC program,[223] we obtain WTMAD2 = 3.58 kcal/mol for SCS-dRPA75-D3BJ (Table 1).

Grimme and Steinmetz proposed[231] the PWPRPA double-hybrid:

$$\begin{aligned}&\mathrm{PWPRPA:}\\ &\mathrm{DH}_{\mathrm{PW6B95,dRPA}}[0,\; 1|1/2, 0.71, 0.35,\; 0.35|0.65\mathrm{NL}(b=10.3)]\end{aligned}$$

The special twist here lies in the use of mGGA orbitals ($c_X=0$), which greatly speeds up the orbital evaluation phase. Hence, with a fast dRPA code, PWPRPA calculations can be a factor of 4–5 faster than double hybrids that include exact exchange during the iterations.

## 9. Summary and Outlook

Double hybrids can, in a sense, be regarded as sitting on the seam line between WTF and DFT methods. For main-group thermochemistry, kinetics, and noncovalent interactions, they can achieve accuracies well beyond those attainable with the best rung four functionals, at a computational cost that is not much greater. They are, arguably, entering the territory of what is attainable with low-cost composite ab initio thermochemical protocols. In addition, they only require a modest number of empirical parameters. Moreover, the superior performance is also seen for types of systems and types of properties *not at all considered* in the parametrization: therefore, such empirical double hybrids do represent more than "fitting round pegs to square holes".

The main Achilles' Heel of the DH approach is static correlation, particularly the Type A variety (absolute near-degeneracy[70]). This means that double hybrids, as currently constituted, are more likely to misbehave for transition metal systems. Possibly this may be mitigated by considering RPA-type methods. Coupling singlet-paired coupled cluster[232] with DFT correlation[233] may represent another path.

Range-separated exchange double hybrids could offer succor in other areas through mitigating self-interaction error and initial applications to optical spectra look very promising. In addition, range-separated correlation might offer[234] an avenue to eliminate or reduce reliance on the dispersion correction at long range.

As a final reflection, we would argue that the choice between purely nonempirical functionals and radically pragmatic approaches with dozens of adjustable parameters like Refs. [29,30,51,56,235] is a false dichotomy. The Berkeley functionals B97M-V,[57] ωB97X-V,[32] and ωB97M-V[33] represent a fruitful middle course, achieving high accuracy and versatility with comparatively modest numbers (around a dozen) of empirical parameters. We have shown that DSD-type double hybrids can in fact achieve even better accuracy with still fewer (4–6) empirical parameters, and that they can reach accuracies comparable to composite wavefunction approaches. The latter either involve similar numbers of parameters (G3, G4) or are much costlier (ccCA, W1, W2-F12), and none of these afford similarly convenient energy derivatives. It therefore appears that empirical double hybrids bring something to the computational modeling table not currently offered by any other approach.

## Acknowledgements

This research was supported by the Israel Science Foundation (grant 1358/15) and by the Minerva Foundation, Munich, Germany, as well as by two internal Weizmann Institute funding sources: the Helen and Martin Kimmel Center for Molecular Design and a research grant from the estate of Emile Mimran. We thank (in alphabetical order of first name) Profs. Amir Karton, Lars Goerigk, Leo Radom, Martin Head-Gordon, Sebastian Kozuch, and Stefan Grimme, as well as Drs. Irena Efremenko, Mark A. Iron, and Narbe Mardirossian, for helpful discussions. Nitai Sylvetsky and Emmanouil Semidalas are thanked for critical reading of the draft manuscript. We also gratefully acknowledge helpful comments by an anonymous reviewer. The artwork in Figure 1 incorporates elements from a public domain photoreproduction of *Le Songe de Jacob* by the French Renaissance painter Nicolas Dipre or d'Ypres (1460–1532).

  

## REVIEW

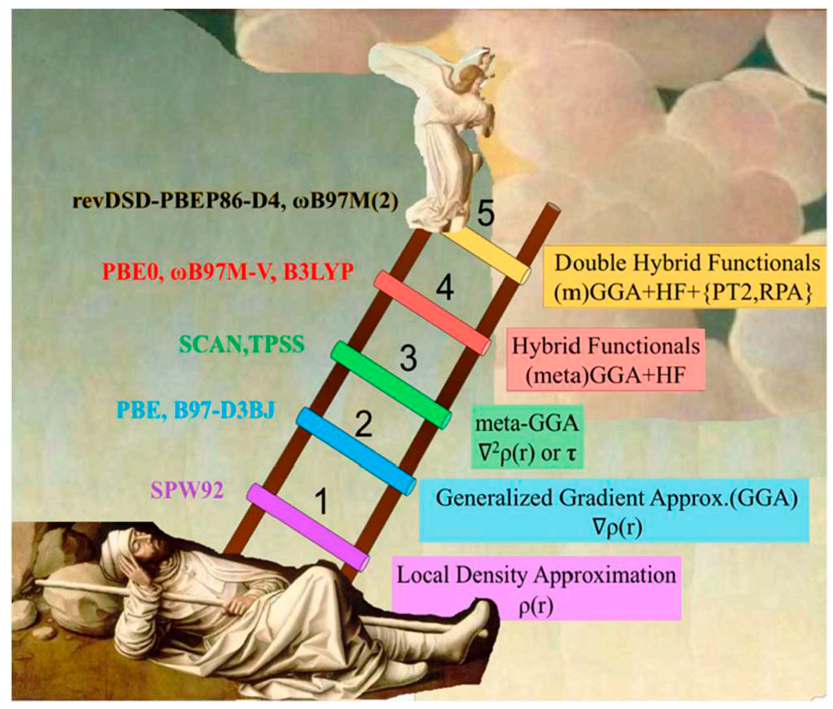

*Prof. Dr. J. M. L. Martin*, G. Santra*

**Empirical Double-Hybrid Density Functional Theory: A 'Third Way' in Between WFT and DFT**